\documentclass[12pt,twoside]{article}   
\usepackage[super,sort,comma]{natbib}
\usepackage[final]{changes}

\usepackage{fancyhdr}		

\usepackage[section]{placeins}   %
\usepackage{amsmath}
\usepackage[superscript]{cite}
\usepackage{caption}
\usepackage{amssymb}

\usepackage{graphicx}

\makeatletter \renewcommand\@biblabel[1]{$^{#1}$} \makeatother
\newcommand{\cen}[1]{\begin{center} #1 \end{center}}

\usepackage[pagewise,mathlines,edtable]{lineno}

\usepackage{hyperref}
\hypersetup{ colorlinks,
    citecolor=blue,
    filecolor=blue,
    linkcolor=blue,
    urlcolor=blue
}

\usepackage{xcolor}

\definecolor{gray}{rgb}{0.6,0.6,0.6}
\definecolor{red}{rgb}{0.85,0,0}
\definecolor{green}{rgb}{0,0.85,0}
\definecolor{blue}{rgb}{0,0,0.85}
\definecolor{beige}{rgb}{0.92,0.87,0.78}
\usepackage[all]{hypcap}    

\begin{document}

\cen{\sf {\Large {\bfseries \replaced{A Language Vision Model Approach for Automated Tumor Contouring in Radiation Oncology}{Advancing Global Health Equity through Oncology Contouring Copilot: A Language Vision Model Approach for Automated Tumor Contouring} } \\  
\vspace*{10mm}
\author{Yi Luo$^{1}$ \and 
        Hamed Hooshangnejad$^{1,2}$ \and 
        Xue Feng$^{3}$ \and 
        Gaofeng Huang$^{3}$ \and 
        Xiaojian Chen$^{1}$ \and 
        Rui Zhang$^{4}$ \and 
        Quan Chen$^{5}$ \and 
        Wil Ngwa$^{2}$ \and 
        Kai Ding$^{2}$}
Yi Luo{$^{1}$} $|$ Hamed Hooshangnejad{$^{1, 2}$} $|$ Xue Feng{$^{3}$} $|$ Gaofeng Huang{$^{3}$} $|$ Xiaojian Chen{$^{1}$} $|$ Rui Zhang{$^{4}$} $|$ Quan Chen{$^{5}$} $|$ Wil Ngwa{$^{2}$} $|$ Kai Ding{$^{2}$}} \\
 $^{1}$ Department of Biomedical Engineering, Johns Hopkins University, Baltimore, MD 21287, USA\\
 $^{2}$ Department of Radiation Oncology and Molecular Radiation Sciences, Johns Hopkins University, Baltimore, MD 21287, USA \\
 $^{3}$ Carina Medical LLC, KY 40513, USA \\
 $^{4}$ Division of Computational Health Sciences, Department of Surgery, University of Minnesota, Minneapolis, MN 55455, USA \\
 $^{5}$ Department of Radiation Oncology, Mayo Clinic Arizona, Phoenix AZ, 85054, USA
\vspace{5mm}\\
}

\def\thefootnote{}%
\footnotetext{Yi Luo and Hamed Hooshangnejad should be considered joint first author.}

\pagenumbering{roman}
\setcounter{page}{1}
\pagestyle{plain}
\noindent Correspondence\\
Kai Ding, Department of Radiation Oncology and Molecular Radiation Sciences, Johns Hopkins University, Kimmel Comprehensive Cancer Center, 401 N Broadway, Suite 1440, Baltimore, MD 21231-2410, USA. Email: kai@jhu.edu\\


\begin{abstract}

\noindent {\bf Background:} \replaced{Lung cancer ranks as the leading cause of cancer-related mortality worldwide}{Global health disparities, especially in oncology within low- and middle-income countries (LMICs), hinder effective non-small cell lung cancer (NSCLC) management}. The complexity of tumor delineation, crucial for radiation therapy, requires expertise often unavailable in resource-limited settings. Artificial Intelligence (AI), particularly with advancements in deep learning(DL) and natural language processing (NLP), offers potential solutions yet is challenged by high false positive rates.\\
{\bf Purpose:} The Oncology Contouring Copilot (OCC) system is developed to leverage oncologist expertise for precise tumor contouring using textual descriptions, aiming to \replaced{increase the efficiency of oncological workflows by combining the strengths of AI with human oversight.}{bridge the oncological care gap in LMICs and improve global health equity.}\\
{\bf Methods:} Our OCC system initially identifies nodule candidates from CT scans. Employing Language Vision Models (LVMs) like GPT-4V, OCC then effectively reduces false positives with clinical descriptive texts, merging textual and visual data to automate tumor delineation, designed to elevate the quality of oncology care \replaced{by incorporating knowledge from experienced domain experts.}{in resource-limited environments with knowledge from experienced domain experts.}\\
{\bf Results:} Deployments of the OCC system resulted in a significant reduction in the false discovery rate by 35.0\%, a 72.4\% decrease in false positives per scan, and an F1-score of 0.652 across our dataset for unbiased evaluation.\\
{\bf Conclusions:} OCC represents a significant advance in oncology care, particularly \replaced{through the use of the latest LVMs to improve contouring results by}{in promoting global health equity by} (1) \replaced{streamlining oncology treatment workflows by optimizing tumor delineation, reducing manual processes}{democratizing access to advanced cancer treatment technologies in LMICs}; (2) offering a scalable and intuitive framework to reduce false positives in radiotherapy planning using LVMs; (3) introducing novel medical language vision prompt techniques to minimize LVMs hallucinations with ablation study, and (4) conducting a comparative analysis of LVMs, highlighting their potential in addressing medical language vision challenges. \\

\end{abstract}

\newpage     

\tableofcontents

\newpage

\setlength{\baselineskip}{0.7cm}      

\pagenumbering{arabic}
\setcounter{page}{1}
\pagestyle{fancy}

\section{Introduction}
Lung cancer is the leading cause of cancer-related deaths worldwide~\cite{leiter2023global}. It is predominantly diagnosed as non-small cell lung cancer (NSCLC)~\cite{bodor2020biomarkers}, for which stereotactic body radiotherapy is often the preferred treatment method when it is inoperable and early-stage. The efficacy of such therapy is contingent upon the quality of radiotherapy treatment planning, which is central to the accurate and laborious localization of lung tumor targets~\cite{zhang2024generalizable}. \deleted{In recent years, analyses of inequities related to cancer therapeutics have shown important disparities across countries and various racial groups~\cite {ringborg2024strategies, ngwa2021role}. These disparities are acute in LMICs, characterized by a scarcity of medical infrastructure and a severe shortage of specialized healthcare professionals.} In developing nations, millions are deprived of access to radiation therapy (RT), \replaced{prompting a growing need for automatic oncology delineation technologies.}{prompting a growing number of radiation oncology professionals to engage in global health initiatives in radiation oncology~\cite{ngwa2016emerging}.} Information and communication technologies (ICTs) hold immense promise in \replaced{greater space-time flexible collaborative action against cancers.}{fostering global health collaborations}~\cite{ngwa2016closing}. The deployment of ICTs, encompassing social media platforms, websites, voice-over messaging, and toll-free telecommunication services, is steadily expanding in the realm of oncology services~\cite{addai2021covid}. Nonetheless, the implementation of remote tumor delineation poses significant time-related challenges, emphasizing the critical need for the development of automated methods for tumor localization.

The considerable advances in deep learning have substantially redirected research efforts towards utilizing these methods for precise localization of lung tumors, predominantly through the analysis of CT volumes~\cite{ding2017accurate, tang2018automated, zhu2018deeplung}. However, accurately distinguishing false positives caused by pulmonary blood vessels,  lung borders, and CT scan noise continues to be a significant challenge~\cite{yu2023deep}. Prior studies aimed at minimizing false positives in lung nodules have largely concentrated on single-modality vision inputs~\cite{setio2016pulmonary, dou2016multilevel, xie2018knowledge}. These studies often overlook the invaluable textual information provided by diagnostic physicians, including radiologists and pathologists, through Electronic Health Records (EHRs). This textual data, abundant in comprehensive pathology and radiology reports, represents a crucial resource for enhancing accurate tumor delineation and reducing false positives.

Recent advancements have seamlessly integrated visual modalities with LLMs, giving rise to LVMs such as GPT-4V~\cite{openai2023gpt} and Claude3 for application in visual commonsense reasoning~\cite{huang2023language}, visual question answering, and multimodal dialogue systems~\cite{alayrac2022flamingo, wu2023visual, openai2023gpt} and so on. While recent studies have shown that GPT-4V is proficient in distinguishing between various medical imaging modalities and anatomical structures, it encounters considerable difficulties when dealing with complex medical issues in detail~\cite{wu2023can}, and often suffers from hallucinations when detecting small-scale objects. 

\begin{figure}[t]
    \begin{center} 
        \includegraphics[width=\linewidth]{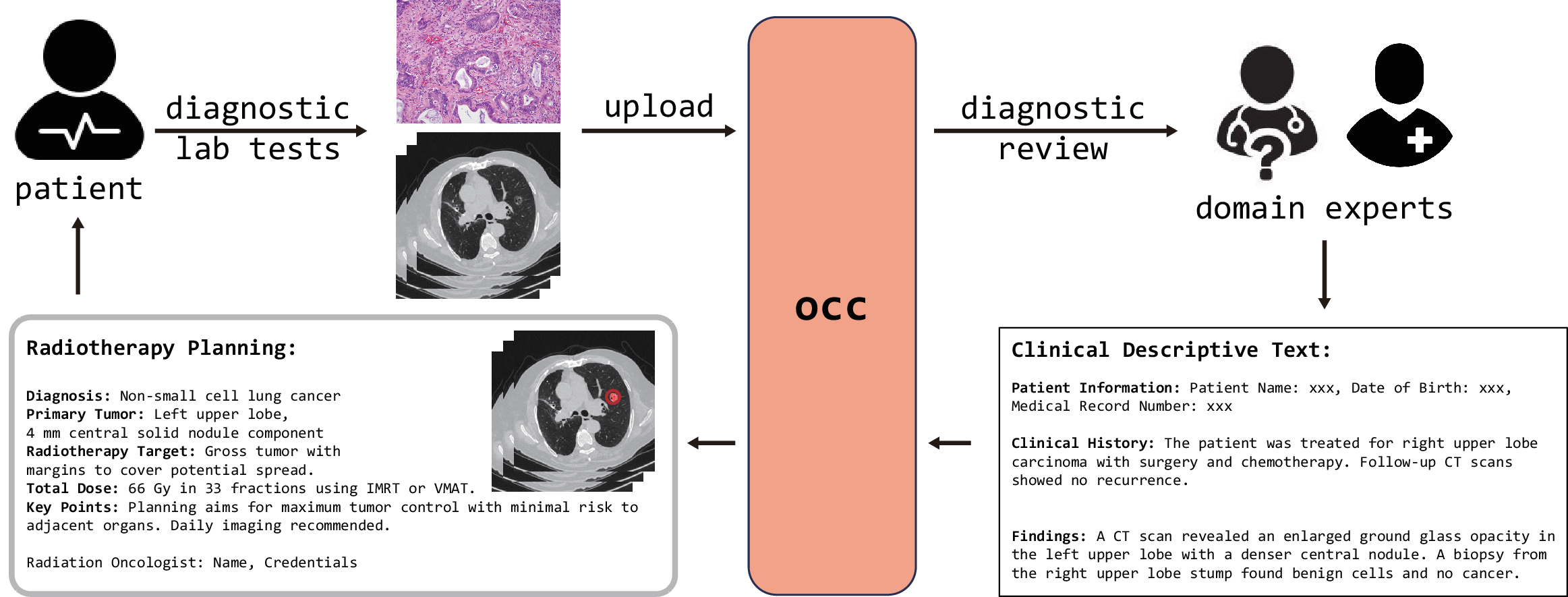} 
        \caption{OCC workflow \deleted{for LMIC patients}: Individuals \deleted{in LMICs} initially undergo diagnostic lab tests including CT scans and pathology biopsies, which are then uploaded to our OCC system. \replaced{Remote domain experts}{Domain experts} review these scans and compile a clinical description of the findings. This narrative, together with the original CT images, is subsequently uploaded to the OCC system. Utilizing this comprehensive data, the system precisely contours \replaced{true positive}{TP} nodules, facilitating targeted and effective radiotherapy planning.
        \label{fig:workflow}
        }
    \end{center}
\end{figure}

Recognizing the limitations of current deep learning approaches. the emerging of powerful LVMs, and the untapped potential of textual data, our work introduces the OCC system as shown in \replaced{Figure}{figure} \ref{fig:workflow}. This system is engineered to integrate the textual descriptions with the visual data from CT scans, leveraging the sophisticated capabilities of LVMs to enhance tumor delineation accuracy. \added{In descriptive text analysis, personalized preferences and varying levels of expertise among human evaluators can introduce biases and inconsistencies, such as differences in expression and terminology. However, the OCC system emphasizes positional information extraction to minimize false positives. Thanks to the advanced capabilities of the latest LVMs, the OCC system can still accurately interpret content and effectively match candidate nodules with the corresponding clinical descriptive texts, despite variations in terminology and phrasing.}Ultimately, the OCC system enables patients \deleted{in LMICs} to benefit from experienced domain experts who can remotely analyze CT scans along with pathology slices\replaced{. By accurately delineating tumor contours based on straightforward clinical text descriptions, the system enhances the precision of radiotherapy planning. This approach not only reduces the reliance on on-site experts but also provides an innovative solution for delivering high-quality, personalized care to patients in resource-limited settings, ultimately improving treatment outcomes.}{, and accurately delineates tumor contours using straightforward clinical text descriptions for better radiotherapy planning, thus improving medical equity, closing the gap in global health disparities, offering a vital step forward in the fight against cancer worldwide.}
\\
\section{Related Work}
Lung nodule segmentation is mainly involved with two steps: candidate nodule detection and false positive reduction. (i) Candidate Nodule Detection: Various methods have been proposed such as Faster Region-based Convolutional Neural Network (R-CNN)~\cite{ding2017accurate}, 3D R-CNN~\cite{tang2018automated}, and faster R-CNN architecture by incorporating dual region proposal networks and a deconvolutional layer~\cite{xie2019automated}. (ii) False Positive Reduction: Research teams have proposed a multi-view ConvNet approach~\cite{setio2016pulmonary},  a CNN model with hand-crafted features~\cite{teramoto2016automated}. Most recently, research primarily focuses on enhancing 3D CNN architectures with attentive 3D-CNN module~\cite{zhao2023attentive}, 3D IRes2Net module~\cite{liu2023d}, and 3D cuboid attention module~\cite{wang2023controlling}. At the same time, notable contributions, e.g., Hooshangnejad et al. ~\cite{hooshangnejad2024exact} integrated EHR information, offered a novel approach to lung nodule false positive reduction. 

Prompt Engineering: Prompt engineering is a method to
 adapt a large pre-trained model to a downstream problem with task-specific hints. Prompts can be divided into two main categories: hard prompts and soft prompts. Hard prompts are manually crafted text prompts with discrete tokens, and soft prompts, are
 optimizable, learnable tensors concatenated with input embeddings, but lack human readability due to their non-alignment with real word embeddings\cite{gu2023systematic}. Due to the limited amount of clinical data and the necessity for user-friendly interfaces, hard prompts are frequently employed in prompt engineering for clinical applications. Recent developments in Language Vision Models (LVMs), as highlighted by studies\cite{alayrac2022flamingo,wu2023can}, have garnered interest in the field of Language Vision prompt engineering. Despite of this growing attention, the practical application of LVMs to medical issues remains
 relatively unexplored.

\section{Methods}
\begin{figure}[t]
    \begin{center}
        \includegraphics[width=\linewidth]{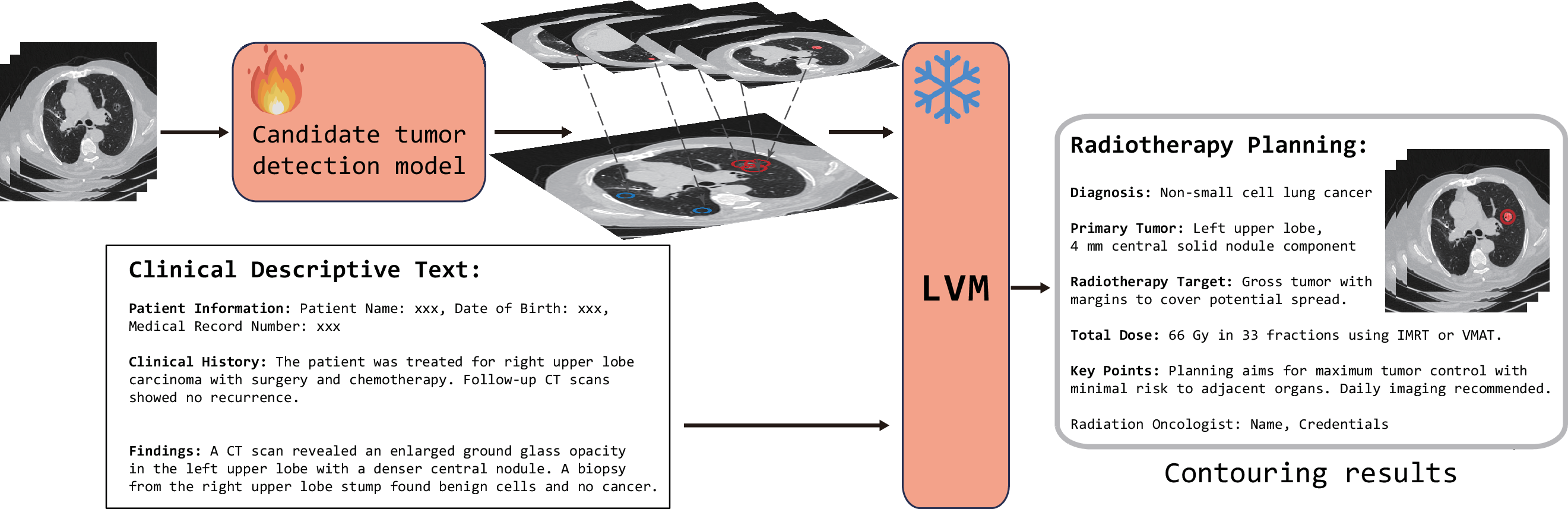} 
        \caption{Components of the OCC model,  the Candidate Tumor Detection component is fully trainable (heated), while the LVM, which serves as the False Positive Reduction model is frozen.}
        \label{fig:occ_model} 
    \end{center} 
\end{figure}

Our OCC system comprises two primary components as shown in \replaced{Figure}{figure} \ref{fig:occ_model}. The first component is the candidate tumor detection model, which uses CT scans from patients to identify multiple potential tumor nodules. The second component is the false positive reduction method. Unlike conventional deep learning-based systems for false positive elimination, our approach utilizes LVMs, which take both the nodule candidates identified in the first component and the domain experts' clinical text descriptions as inputs. By processing and understanding these descriptions, the LVMs effectively select the correct tumor nodules, thereby enhancing the accuracy of false positive removal. Ultimately, the OCC system empowers patients \deleted{in LMICs}to achieve precise lung nodule delineation under the guidance or second opinion of remote, experienced domain experts in high-patient volume centers, facilitating radiotherapy planning for patients \deleted{in LMICs}. 

\subsection{Candidate Tumor Detection Model}
\subsubsection{Architecture}
We deployed the Retina-UNet3D~\cite{hooshangnejad2024exact} as our candidate tumor detection model. Retina-UNet3D aligns the principle of both feature pyramid network (FPN)~\cite{lin2017focal} and UNet~\cite{ronneberger2015u}, which enables the model to harness the advantages of FPN within 3D detection framework, while also building upon the proven effectiveness of UNet-3D for segmentation tasks. \added{The detailed architecture of the Retina-UNet3D model is presented in Supplementary Figure 1.}

\subsubsection{Loss Functions}
We used a dual loss by adding Categorical Cross-Entropy Loss $L_c$ and Dice Loss $L_d$.
By combining them into a unified dual loss function, this strategy harmonized pixel-level precision with image-level authenticity, enhancing the segmentation process. It bolstered the model's capacity for recognizing intricate object details and simultaneously advanced a thorough evaluation of image quality. 
\[
L_c\left(p,q\right)=-\sum_iy_i\text{log}\left(p_i\right) \tag{1}
\]
\[
L_d\left(y,p\right)=1-\frac{2\sum_iy_ip_i}{\sum_iy_i^2+\sum_ip_i^2} \tag{2}
\]
\[
L=L_c+L_d \tag{3}
\]
We adopted smooth L1 loss~\cite{armato2011lung} and focal loss~\cite{ren2015faster} for box regression heads and box classification heads separately. 
\[
\left.smoothL1=\left\{\begin{array}{cc}\frac{0.5(MAE)^2}{\delta}&ifMAE<1\\MAE-0.5\delta&otherwise\end{array}\right.\right. \tag{4}
\]
\[
\left.F_L\left(p,y\right)=\left\{\begin{array}{cc}-\alpha(1-p)^{\gamma}\log\left(p\right)&ify=1\\-\left(1-\alpha\right)p^{\gamma}\log\left(1-p\right)&ify=0\end{array}\right.\right. \tag{5}
\]
\subsection{False Positive Reduction Model}
We mainly integrated GPT-4V as the cornerstone of our model to process both visual and linguistic data concurrently. Following the acquisition of segmentation outputs from our Candidate Tumor Detection Model, we select a random slice that includes representations of nodule and lobe masks. This image, along with the patient's Electronic Health Record (EHR) which provides clinical descriptions, is fed into GPT-4V. GPT-4V then autonomously identifies the location of potential nodules, correlating them with definitive clinical diagnoses extracted from the EHR. In its final phase, GPT-4V generates a concise and clear textual report, helping effectively minimize false positives.

\subsubsection{Experiment Design}
In our research, we combined image and text prompts with LVMs, aligning their outputs with oncologist assessments. We identified a highly effective prompt engineering approach and delved into its potential, yielding six key insights for enhancing LVMs use in medical contexts with an ablation study, conducted from November 2023 to March 2024. Additionally, we implemented a UNet-3D-based false positive reduction network, using it as a comparison for our LVM-based Model.

\subsubsection{Medical Language Vision Prompt Methods}
\begin{figure}[t]
    \begin{center}
        \includegraphics[width=\linewidth]{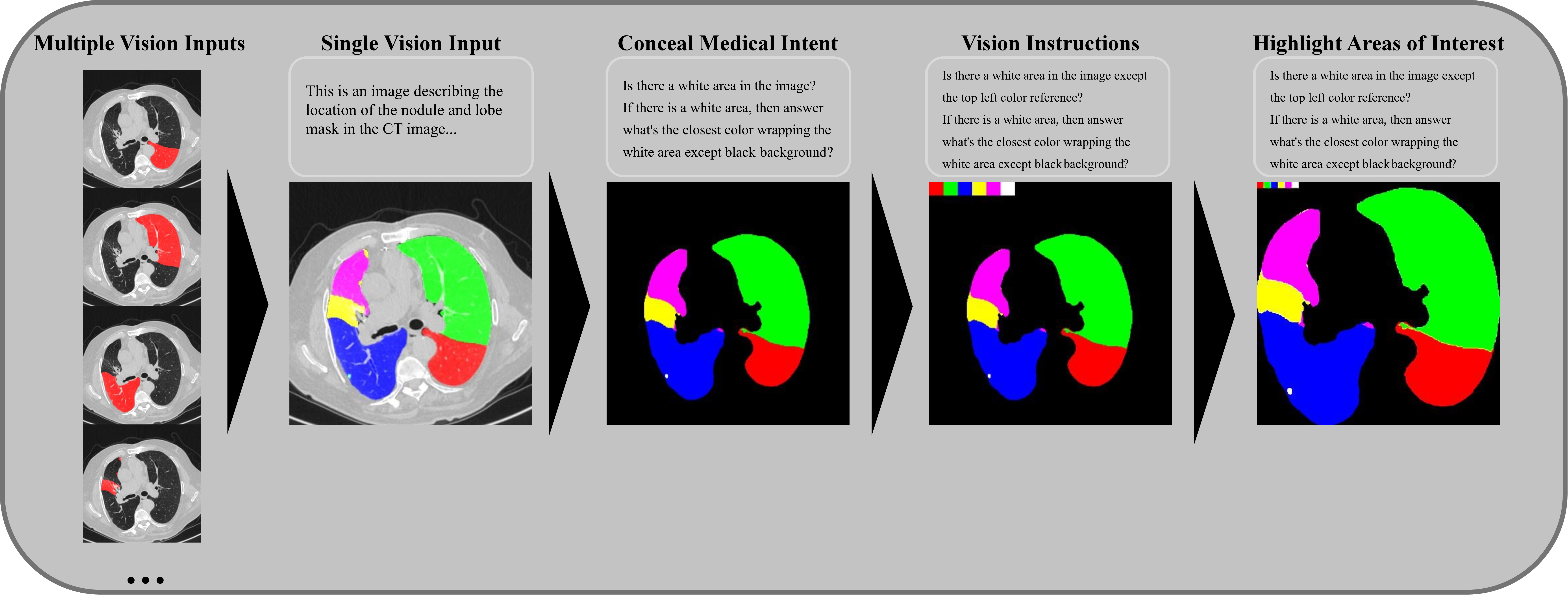} 
        \caption{\added{The figure illustrates several medical vision language prompt methods. To simplify analysis, masks were combined into a single image, transitioning from a single vision input to multiple vision inputs. To conceal medical intent, the CT chest wall background was removed. A color reference was added in the top left corner for vision instructions, and contrast was enhanced while the marginal background was removed to highlight areas of interest.}}
        \label{fig:prompt} 
    \end{center} 
\end{figure}

\textbf{Single Vision Input} 
Employing a single image to present all spatial information relevant to the nodule and lung lobe masks outperformed the use of multiple images, thereby simplifying the visual input for the model.
\textbf{Leave Time to Think}
We gave the model sufficient time to process information, avoiding word limits that could truncate its reasoning process. This approach led to richer, more in-depth, and more accurate responses.
\textbf{Conceal Medical Intent}
To enhance the accuracy of GPT-4V's outcomes, we rephrased medical prompts into a generalized language, which allowed us to circumvent the AI's default restrictive responses.
\textbf{A Series of Guiding Questions} 
We broke down complex medical queries into simpler questions, which directed the AI through a logical reasoning process and resulted in more consistent and correct answers.
\textbf{Vision Instructions} 
By embedding color references in the images, we significantly improved GPT-4V’s color recognition capabilities, aiding its performance in identifying objects with various colors.
\textbf{Highlighting Areas of Interest}
To improve the AI's ability to detect small nodal areas, we cropped out extraneous backgrounds and adjusted the image contrast, making crucial details more discernible.
\added{An example of our several medical language vision prompt methods is shown in Figure \ref{fig:prompt}.}

\subsection{Dataset and Preprocessing}
In the development of our candidate nodule detection model, we utilized the Lung Image Database Consortium imaging collection (LIDC-IDRI) from the Cancer Imaging Archive (TCIA)~\cite{tang2019automatic}, renowned for its high-quality annotations of lung nodules. Given the lack of transparency regarding the datasets used to train the currently popular commercial LVMs, we aimed to develop and validate our false positive reduction model without bias. To achieve this, we utilized 31 CT datasets including 10 diagnostic CT datasets and 21 planning CT datasets obtained from the stereotactic body radiation therapy patients treated at Johns Hopkins Hospital. Importantly, only one diagnostic CT data was utilized for development, others were used for validation as unseen data. To safeguard privacy, patient-specific details are thoroughly annotated by manual. Additionally, intending to minimize the randomness of the responses, we set the temperature parameter at zero during the experimental trials to make it fully reproducible.

\subsection{Evaluation Metrics}
We utilize universally recognized nomenclature. A \replaced{True Positive(TP)}{$TP$} denotes an instance where the model correctly identifies a verified lung nodule, thus indicating a precise prediction. Conversely, a \replaced{False Positive(FP)}{$FP$} arises when the model erroneously identifies a non-nodule as a nodule \added{. A False Negative (FN) arises when the model fails to detect an actual nodule, leading to a missed detection, while a True Negative (TN) arises when the model correctly identifies an area as non-nodule, confirming an accurate negative prediction}. The aggregate count of lung nodules within the dataset is represented by \( N_{\text{sample}} \). \( N_{\text{reject}} \) means the total amount of our requests rejected by internal inhibition of LVMs. We deployed the False Discovery Rate(FDR), reject rate, \replaced{FP/scan, Sensitivity(Sen), Specificity(Spe), and F1-score}{and FP/Scan} to measure our model.
\[
FDR=\frac{FP}{TP+FP} \tag{6}
\]
\added{
\[
Sensitivity=\frac{TP}{TP+FN} \tag{7}
\]
\[
Specificity=\frac{TN}{TN+FP} \tag{8}
\]
\[
\textit{F1-score} = \frac{2 \times TP}{2 \times TP + FP + FN} \tag{9}
\]
}
\[
\textit{Reject Rate}=\frac{N_{reject}}{N_{sample}} \tag{\replaced{10}{7}}
\]
\[
FP/Scan=\frac{FP}{N_{sample}} \tag{\replaced{11}{8}}
\]

\section{Results}
In our experiment, the training dataset comprised a single diagnostic CT scan featuring 7 candidate nodules. For validation purposes, the dataset included 9 diagnostic CT scans and 21 planning CT scans, collectively presenting 221 candidate nodules, there are 2 cases with no candidate nodules detected, so we excluded them in our following analysis. \added{Then, }We conducted a comprehensive evaluation of our OCC system with different false positive reduction cores including the UNet-3D-based method, and several LVM-based methods, specifically, 
ViLT~\cite{kim2021vilt}, Claude3 Sonnet, and GPT-4V, all tested with the same well-designed prompt inputs. The outcomes of this analysis are presented in Table \ref{tab:method_comparison}. \added{The probability density function curves for FP/Scan with various False Positive reduction methods are illustrated in Supplementary Figure 2.}
\\
\begin{table}[ht]
\centering
\caption{Comparison of different FP reduction methods.}
\begin{tabular}{lccccc}
\hline
\textbf{Method} & \textbf{FDR}\textsuperscript{\(\downarrow\)}& \textbf{Average FP/Scan}\textsuperscript{\(\downarrow\)} & \textbf{Sen}\textsuperscript{\(\uparrow\)} & \textbf{Spe}\textsuperscript{\(\uparrow\)} & \textbf{F1-score}\textsuperscript{\(\uparrow\)}\\
\hline
Candidates & 0.787 & 6.214 & - & - & -\\
Candidates+Unet-3D & 0.696 & 3.357 & 0.872 & 0.460 & 0.366 \\
Candidates+ViLT  & 0.773 & 3.036 & 0.532 & 0.511 & 0.318 \\
Candidates+Claude3 Sonnet  & 0.556 & 1.964 & 0.936 & 0.684 & 0.603 \\
Candidates+GPT-4V\textbf{(Ours)} & 0.511 & 1.714  & 0.979 & 0.724 & 0.652 \\
\hline
\end{tabular}
\label{tab:method_comparison}
\end{table}
\\
\begin{table}[ht]
\centering
\caption{Ablation study results about impacts of medical language vision prompt methods.}
\begin{tabular}{lccccccc}
\hline
\multicolumn{1}{c}{Methods and Metrics} & \multicolumn{7}{c}{Choice and Results} \\
\hline
Single Vision Input & $\checkmark$ & $\checkmark$ & $\checkmark$ & $\checkmark$ & $\checkmark$ & & $\checkmark$\\
Leave Time to Think & $\checkmark$ & $\checkmark$ & $\checkmark$ & $\checkmark$ & & $\checkmark$ & $\checkmark$\\
Conceal Medical Intent & $\checkmark$ & $\checkmark$ & $\checkmark$ & & $\checkmark$ & $\checkmark$ & $\checkmark$\\
A Series of Guiding Questions & $\checkmark$ & $\checkmark$ & & $\checkmark$ & $\checkmark$ & $\checkmark$ & $\checkmark$\\
Vision Instructions & $\checkmark$ & & $\checkmark$ & $\checkmark$ & $\checkmark$ & $\checkmark$ & $\checkmark$\\
Highlighting Areas of Interest &  & $\checkmark$ & $\checkmark$ & $\checkmark$ & $\checkmark$ & $\checkmark$ & $\checkmark$\\ 

\hline
FDR\textsuperscript{\(\downarrow\)}      & 0.615 & 0.546 & 0.667 & 0.715 & 0.639 & 0.762 & 0.511 \\
Average FP/Scan\textsuperscript{\(\downarrow\)} & 2.286 & 1.893 & 2.607 & 4.036 & 1.393 & 1.607 & 1.714 \\
Sen\textsuperscript{\(\uparrow\)} & 0.833 & 0.917 & 0.766 & 0.954 & 0.468 & 0.298 & 0.979 \\
Spe\textsuperscript{\(\uparrow\)} & 0.630 & 0.695 & 0.580 & 0.362 & 0.776 & 0.741 & 0.724 \\
F1-score\textsuperscript{\(\uparrow\)} & 0.527 & 0.607 & 0.464 & 0.439 & 0.407 & 0.265 & 0.652\\
Reject Rate\textsuperscript{\(\downarrow\)} & - & - & - & 0.575& -& -& -\\

\hline
\end{tabular}
\label{tab:ablation_study}
\end{table}

An optimal model ought to sustain elevated sensitivity, ensuring that true positive nodules are not erroneously classified as false negatives. Concurrently, it should endeavor to minimize the FDR and the average FP per scan, as well as the reject rate, while enhancing measures of sensitivity, specificity, and the F1 score, to achieve a balanced diagnostic accuracy. Our proposed workflow, which incorporates state-of-the-art LVM models such as Claude3 Sonnet and GPT-4V, demonstrates substantial enhancements in various metrics over traditional deep learning-based false positive reduction methods. When incorporating the ViLT model as the core component of our LVM framework, we encounter a significant limitation due to the model's inherent capabilities. Given a segmentation map, the model frequently hallucinates, mistakenly reclassifying true positive nodules as false positives. This error leads to an obvious decrease in performance.

Moreover, we executed an ablation study to assess the efficacy of the medical language vision prompt methods we devised. In these trials, we systematically removed individual elements from our prompt engineering strategies, ensuring that the remaining methods were still executed. \added{Every column displays the results of the test sets based on various combinations of medical prompt engineering strategies.} The results of these tests are shown in Table~\ref{tab:ablation_study}.

The results of our ablation study indicate that the absence of any element within our prompt engineering methodology significantly affects model performance. The omission of "Single Vision Input" highlights the LVM's limited capacity for comparative analysis across multiple image inputs. When "Time to Think" is removed, the LVM tends to bypass preliminary contemplation and leaps to conclusions without logical validation, disrupting the natural flow of reasoned thinking. The lack of "Conceal Medical Intent" often causes the LVM to internally filter out such medical inquiries, leading to an increased presence of false positive nodules that are not properly dismissed. Without "A Series of Guiding Questions," the LVM's logical reasoning occasionally falters. The absence of "Vision Instructions" can cause the LVM's inherent color biases to skew its judgments. Finally, without "Highlighting Areas of Interest," the model struggles to pinpoint the exact location within a vast background, which can impede accurate classification.

\section{Case Study}
We conducted a case study to assess the false positive reduction capability of different LVMs within the OCC system for nodule images. This study was designed with two goals, i.e., whether the models could correctly identify TP or FP nodules (G1), and the richness of the information inferred by the models (G2). Summarily, we input various nodule images and the questions into the GPT-4V, Claude3 Sonnet, and ViLT. We highlighted the key responses in red for correct inferences, and in blue for incorrect ones.
\\
\begin{figure}[h]
    \begin{center} 
        \includegraphics[width=\linewidth]{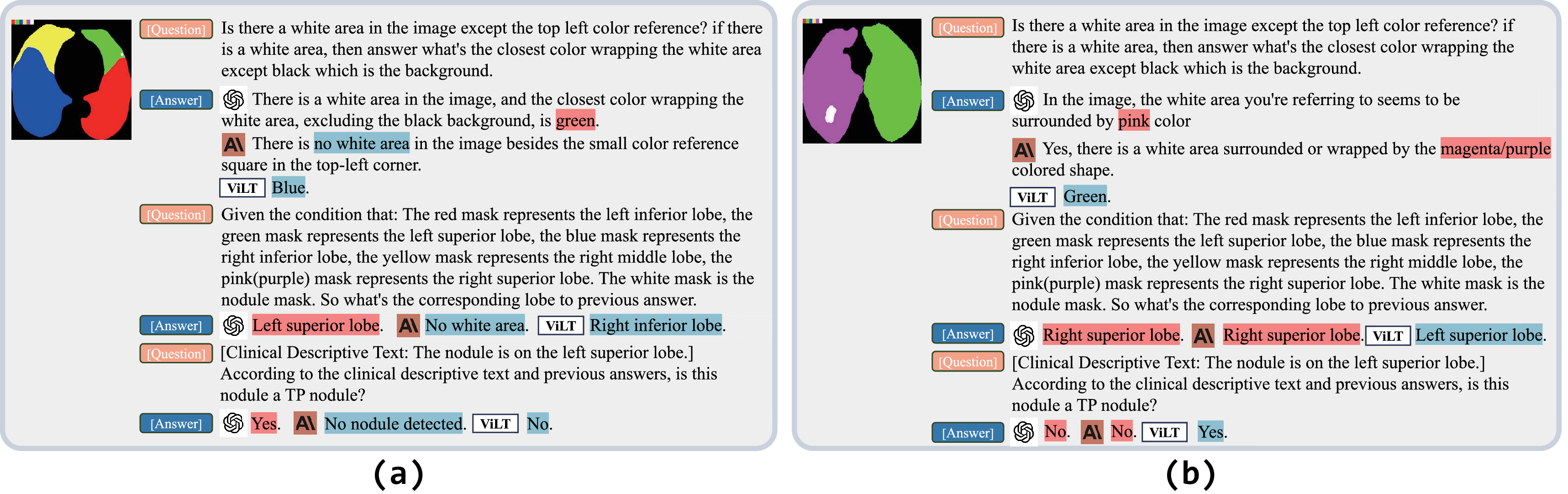} 
        \caption{Cases of false positive reduction in the OCC system with different LVMs. (\textbf{a}) GPT-4V model accurately identifies the small TP nodule, whereas Claude3 Sonnet provides no response, and ViLT erroneously classifies it as an FP.
        (\textbf{b}) ViLT mistakenly labels the nodule as a TP, while both GPT-4V and Claude3 Sonnet correctly identify and eliminate the FP nodule.
        \label{fig:user-study}
        }
    \end{center}
\end{figure}
The first case is conducted for a TP scenario, which is a small nodule data on the left upper lobe. As shown in Figure~\ref{fig:user-study} (a), the GPT-4V generally could give the right answer, demonstrating its ability to identify the TP nodule (G1). However, the performance of Claude3 Sonnet and ViLT was subpar, indicating their poor performance on small nodule detection. In addition, the answer from GPT-4V and Claude3 Sonnet is a paragraph, while the ViLT only inputs a simple sentence or even a single word (G2). 

The results of the second case are shown in Figure~\ref{fig:user-study} (b), where both GPT-4V and Claude3 Sonnet made the correct answers for a medium FP nodule successfully. On the other hand, the ViLT responded the wrong answer with few words.


\section{Discussion}
In our study, we introduced a novel OCC system to help radiotherapy planning for patients \deleted{in LMICs}, leveraging the capabilities of LVMs. This novel method synergistically integrates visual data from clinical CT scans with textual information from experienced domain experts, culminating in a highly effective reduction of false positives, enhancing the precision of radiotherapy treatment. Furthermore, our approach not only capitalizes on the advanced features of LVMs but also holds the potential for widespread adoption across various clinical settings.

Our research is subject to certain limitations. Firstly, the false positive reduction and language vision prompt engineering strategies we introduced were tested exclusively within our institution's dataset\added{, which was provided by leading radiation oncologists and included precise tumor delineations along with comprehensive clinical descriptive texts}. While the results were promising, it is imperative to conduct further evaluations on a more extensive and varied dataset. \added{When working with larger datasets, obtaining evaluations from top radiation oncologists is challenging. Additionally, comprehensive and reliable human evaluations often require a significant number of evaluators, making the process time-consuming, labor-intensive, and costly. In this context, LLM-as-a-Judge \cite{zheng2023judging} could provide a potential solution to get accurate and detailed clinical descriptive text datasets.}\replaced{Besides}{Secondly}, \added{Although the FDR in our experiment has been reduced, it remains relatively high at 0.511. Additionally, the 3D DICE score for TP nodules across the experiments is 0.4164, indicating that the performance is still not suitable for real-world clinical applications. This is because} our \added{candidate tumor detection model prioritizes sensitivity to ensure that no nodules are missed to meet the clinical needs. However, this approach results in a significant number of low-quality nodule predictions. Besides, our false positive reduction} methodology predominantly mitigates false positives by correlating the candidate nodule's location with the clinical descriptive texts. However, for false positive nodules situated within the target lobe, our approach faces challenges in excluding these erroneous detections, Further exploration into the use of varied textual information \added{such as tumor size and tumor stage} for nodule contouring holds significant promise. \added{Additionally, our medical vision-language prompt methods primarily aim to enhance visual focus to reduce hallucinations when detecting small-scale objects. However, beyond small-scale object hallucinations, there are other types of hallucinations, such as positional, verbosity, and self-enhancement biases\cite{zheng2023judging}. Further research is needed to explore solutions for mitigating these hallucinations in medical applications.} 

\section{Conclusion}
We introduce the OCC system, which integrates LVMs with clinical CT imaging and textual data, \replaced{to reduce the need for on-site experts while ensuring precise and efficient radiation therapy.}{representing a potential advancement in addressing oncological care disparities in LMICs.} The system enhances the reliability of clinical applications by implementing novel medical language vision prompt techniques that effectively reduce hallucinatory outputs from LVMs. Moreover, we provide a scalable and intuitive framework that significantly improves diagnostic accuracy and the quality of oncology care in resource-limited environments. A detailed comparative analysis of LVMs within the OCC highlights the LVMs' transformative potential for medical language vision problems.

\clearpage

\section*{References}
\addcontentsline{toc}{section}{\numberline{}References}
\vspace*{-10mm}


\begin{thebibliography}{10}

\bibitem{leiter2023global}
A.~Leiter, R.~R. Veluswamy, and J.~P. Wisnivesky,
\newblock The global burden of lung cancer: current status and future trends,
\newblock Nature Reviews Clinical Oncology {\bf 20}, 624--639 (2023).

\bibitem{bodor2020biomarkers}
J.~N. Bodor, Y.~Boumber, and H.~Borghaei,
\newblock Biomarkers for immune checkpoint inhibition in non--small cell lung cancer (NSCLC),
\newblock Cancer {\bf 126}, 260--270 (2020).

\bibitem{zhang2024generalizable}
L.~Zhang et~al.,
\newblock Generalizable and promptable artificial intelligence model to augment clinical delineation in radiation oncology,
\newblock Medical Physics  (2024).

\bibitem{ngwa2016closing}
W.~Ngwa et~al.,
\newblock Closing the cancer divide through Ubuntu: Information and communication technology-powered models for global radiation oncology, 2016.

\bibitem{addai2021covid}
B.~W. Addai and W.~Ngwa,
\newblock COVID-19 and cancer in Africa,
\newblock Science {\bf 371}, 25--27 (2021).

\bibitem{ding2017accurate}
J.~Ding, A.~Li, Z.~Hu, and L.~Wang,
\newblock Accurate pulmonary nodule detection in computed tomography images using deep convolutional neural networks,
\newblock in {\em Medical Image Computing and Computer Assisted Intervention- MICCAI 2017: 20th International Conference, Quebec City, QC, Canada, September 11-13, 2017, Proceedings, Part III 20}, pages 559--567, Springer, 2017.

\bibitem{tang2018automated}
H.~Tang, D.~R. Kim, and X.~Xie,
\newblock Automated pulmonary nodule detection using 3D deep convolutional neural networks,
\newblock in {\em 2018 IEEE 15th international symposium on biomedical imaging (ISBI 2018)}, pages 523--526, IEEE, 2018.

\bibitem{zhu2018deeplung}
W.~Zhu, C.~Liu, W.~Fan, and X.~Xie,
\newblock Deeplung: Deep 3d dual path nets for automated pulmonary nodule detection and classification,
\newblock in {\em 2018 IEEE winter conference on applications of computer vision (WACV)}, pages 673--681, IEEE, 2018.

\bibitem{yu2023deep}
X.~Yu, L.~He, Y.~Wang, Y.~Dong, Y.~Song, Z.~Yuan, Z.~Yan, and W.~Wang,
\newblock A deep learning approach for automatic tumor delineation in stereotactic radiotherapy for non-small cell lung cancer using diagnostic PET-CT and planning CT,
\newblock Frontiers in Oncology {\bf 13} (2023).

\bibitem{setio2016pulmonary}
A.~A.~A. Setio, F.~Ciompi, G.~Litjens, P.~Gerke, C.~Jacobs, S.~J. Van~Riel, M.~M.~W. Wille, M.~Naqibullah, C.~I. S{\'a}nchez, and B.~Van~Ginneken,
\newblock Pulmonary nodule detection in CT images: false positive reduction using multi-view convolutional networks,
\newblock IEEE transactions on medical imaging {\bf 35}, 1160--1169 (2016).

\bibitem{dou2016multilevel}
Q.~Dou, H.~Chen, L.~Yu, J.~Qin, and P.-A. Heng,
\newblock Multilevel contextual 3-D CNNs for false positive reduction in pulmonary nodule detection,
\newblock IEEE Transactions on Biomedical Engineering {\bf 64}, 1558--1567 (2016).

\bibitem{xie2018knowledge}
Y.~Xie, Y.~Xia, J.~Zhang, Y.~Song, D.~Feng, M.~Fulham, and W.~Cai,
\newblock Knowledge-based collaborative deep learning for benign-malignant lung nodule classification on chest CT,
\newblock IEEE transactions on medical imaging {\bf 38}, 991--1004 (2018).

\bibitem{openai2023gpt}
R.~OpenAI,
\newblock Gpt-4 technical report. arxiv 2303.08774,
\newblock View in Article {\bf 2}, 13 (2023).

\bibitem{huang2023language}
S.~Huang et~al.,
\newblock Language is not all you need: Aligning perception with language models,
\newblock arXiv preprint arXiv:2302.14045  (2023).

\bibitem{alayrac2022flamingo}
J.-B. Alayrac et~al.,
\newblock Flamingo: a visual language model for few-shot learning,
\newblock Advances in Neural Information Processing Systems {\bf 35}, 23716--23736 (2022).

\bibitem{wu2023visual}
C.~Wu, S.~Yin, W.~Qi, X.~Wang, Z.~Tang, and N.~Duan,
\newblock Visual chatgpt: Talking, drawing and editing with visual foundation models,
\newblock arXiv preprint arXiv:2303.04671  (2023).

\bibitem{wu2023can}
C.~Wu et~al.,
\newblock Can gpt-4v (ision) serve medical applications? case studies on gpt-4v for multimodal medical diagnosis,
\newblock arXiv preprint arXiv:2310.09909  (2023).

\bibitem{xie2019automated}
H.~Xie, D.~Yang, N.~Sun, Z.~Chen, and Y.~Zhang,
\newblock Automated pulmonary nodule detection in CT images using deep convolutional neural networks,
\newblock Pattern Recognition {\bf 85}, 109--119 (2019).

\bibitem{teramoto2016automated}
A.~Teramoto, H.~Fujita, O.~Yamamuro, and T.~Tamaki,
\newblock Automated detection of pulmonary nodules in PET/CT images: Ensemble false-positive reduction using a convolutional neural network technique,
\newblock Medical physics {\bf 43}, 2821--2827 (2016).

\bibitem{zhao2023attentive}
D.~Zhao, Y.~Liu, H.~Yin, and Z.~Wang,
\newblock An attentive and adaptive 3D CNN for automatic pulmonary nodule detection in CT image,
\newblock Expert Systems with Applications {\bf 211}, 118672 (2023).

\bibitem{liu2023d}
B.~Liu, H.~Song, Q.~Li, Y.~Lin, X.~Weng, Z.~Su, and J.~Yang,
\newblock 3D ARCNN: An Asymmetric Residual CNN for False Positive Reduction in Pulmonary Nodule,
\newblock IEEE Transactions on NanoBioscience  (2023).

\bibitem{wang2023controlling}
S.~Wang, Y.~Liu, and C.~Shi,
\newblock Controlling False-Positives in Automatic Lung Nodule Detection by Adding 3D Cuboid Attention to a Convolutional Neural Network,
\newblock Biomedical Signal Processing and Control {\bf 85}, 104946 (2023).

\bibitem{hooshangnejad2024exact}
H.~Hooshangnejad, X.~Feng, G.~Huang, R.~Zhang, Q.~Chen, and K.~Ding,
\newblock EXACT-Net: EHR-guided lung tumor auto-segmentation for non-small cell lung cancer radiotherapy,
\newblock arXiv preprint arXiv:2402.14099  (2024).

\bibitem{gu2023systematic}
J.~Gu, Z.~Han, S.~Chen, A.~Beirami, B.~He, G.~Zhang, R.~Liao, Y.~Qin, V.~Tresp, and P.~Torr,
\newblock A systematic survey of prompt engineering on vision-language foundation models,
\newblock arXiv preprint arXiv:2307.12980  (2023).

\bibitem{lin2017focal}
T.-Y. Lin, P.~Goyal, R.~Girshick, K.~He, and P.~Doll{\'a}r,
\newblock Focal loss for dense object detection,
\newblock in {\em Proceedings of the IEEE international conference on computer vision}, pages 2980--2988, 2017.

\bibitem{ronneberger2015u}
O.~Ronneberger, P.~Fischer, and T.~Brox,
\newblock U-net: Convolutional networks for biomedical image segmentation,
\newblock in {\em Medical image computing and computer-assisted intervention--MICCAI 2015: 18th international conference, Munich, Germany, October 5-9, 2015, proceedings, part III 18}, pages 234--241, Springer, 2015.

\bibitem{armato2011lung}
S.~G. Armato~III et~al.,
\newblock The lung image database consortium (LIDC) and image database resource initiative (IDRI): a completed reference database of lung nodules on CT scans,
\newblock Medical physics {\bf 38}, 915--931 (2011).

\bibitem{ren2015faster}
S.~Ren, K.~He, R.~Girshick, and J.~Sun,
\newblock Faster r-cnn: Towards real-time object detection with region proposal networks,
\newblock Advances in neural information processing systems {\bf 28} (2015).

\bibitem{tang2019automatic}
H.~Tang, C.~Zhang, and X.~Xie,
\newblock Automatic pulmonary lobe segmentation using deep learning,
\newblock in {\em 2019 IEEE 16th international symposium on biomedical imaging (ISBI 2019)}, pages 1225--1228, IEEE, 2019.

\bibitem{kim2021vilt}
W.~Kim, B.~Son, and I.~Kim,
\newblock Vilt: Vision-and-language transformer without convolution or region supervision,
\newblock in {\em International Conference on Machine Learning}, pages 5583--5594, PMLR, 2021.

\bibitem{zheng2023judging}
L.~Zheng et~al.,
\newblock Judging llm-as-a-judge with mt-bench and chatbot arena,
\newblock Advances in Neural Information Processing Systems {\bf 36}, 46595--46623 (2023).

\end{thebibliography}
\end{document}